\documentclass[prl,twocolumn,showpacs]{revtex4} 
\usepackage{graphicx}
\usepackage{amssymb} 
\usepackage{amsmath}
\usepackage{epstopdf}
\usepackage{bm}
\usepackage{color}
\usepackage{ulem}

\begin{document} 

\title{Modeling photoconversion efficiency of perovskite solar cells}

\author{A.\,V.\,Sachenko$^{1}$, V.\,P.\,Kostylyov$^1$, A.\,V.\,Bobyl$^2$, V.\,M.\,Vlasyuk$^1$, I.\,O.\,Sokolovskyi$^{1,3}$, E.\,I.\,Terukov$^{2,4}$, M.\,Evstigneev$^3$
}

\address{$^1$V. Lashkaryov Institute of Semiconductor Physics, NAS of Ukraine, 03028 Kiev, Ukraine}
\address{$^2$A.F. Ioffe Physical-Technical Institute RAS, 194021 St.-Petersburg, Russia}
\address{$^3$Department of Physics and Physical Oceanography, Memorial University of Newfoundland, St. John's, NL, A1B 3X7 Canada}
\address{$^4$TFTC Ioffe R\&D Center, 194021 St.-Petersburg, Russia}

\begin{abstract}
A theoretical approach to  photoconversion efficiency modeling in perovskite p-i-n structures is developed. The results of this modeling compare favorably with the experiment and indicate that the surfaces of the perovskite solar cells (SCs) are naturally textured. It is shown that photoconversion efficiency in the limiting case of negligible Shockley-Read-Hall and surface recombination and in the absence of optical losses reaches the value of 29\%. In the realistic case, the current-voltage curve ideality factor equals 2. This value is not due to recombination in the space-charge region; rather, it can be explained by taking into account the effect of the rear surface and high excitation level.
\end{abstract}


\maketitle 
Although photoconversion efficiency in perovskite solar cells (SCs) presently achieved is about 20\% \cite{You14, Green14}, there is a lack of publications on the theoretical modeling of these novel energy sources. This makes it difficult to systematically optimize the parameters of perovskite SCs. Here, a theory of photoconversion in p-i-n structures is developed, for which the following criteria are fulfilled: (i) Diffusion length is much higher than the i-region thickness, $L_d \gg d$; (ii) Excess electron-hole pair concentration generated by light notably exceeds the equilibrium carrier concentration in the i-region, $\Delta n \gg n_0$. Note that this inequality holds quite well, because, according to \cite{Han16}, $n_0 \approx 10^{11}$\,cm$^{-3}$ in FAPbI$_3$, whereas the estimated value of $\Delta n$ exceeds $10^{13}$\,cm$^{-3}$. 

These criteria allow one to use the approach introduced earlier for heterojunction solar cells modeling, see \cite{Sachenko15, Sachenko16, Sachenko17}. The distinct feature of this approach is that it accounts for the contribution of the SC’s rear surface, located near the isotype n-n$^+$ junction, both to the current-voltage (I-V) curves and to the open-circuit voltage.

As shown in \cite{Sachenko15, Sachenko16, Sachenko17}, under the conditions (i) and (ii), the light-generated current density at applied bias $V$ is described by the expression
\begin{eqnarray}
&&J_L(V) = J_{SC} + \frac{V - A_{SC} J_LR_S}{A_{SC}R_{sh}} \nonumber \\
&&- J_0\,e^{q(V - A_{SC}J_LR_S)/(2kT)}\ ,
\label{1}
\end{eqnarray}
where $J_{SC}$ is the short-circuit current density, $A_{SC}$ is the SC surface area, $kT$ is thermal energy, $R_S$ and $R_{sh}$ are series and shunt resistance, and $J_0$ is saturation current density. For a p-i-n structure, it is given by 
\begin{equation}
J_0 = q\left(\frac{d}{\tau_{SRH}} + S + dA\Delta n\right)n_i\ .
\label{2a}
\end{equation}
Here, $q$ is the elementary charge, $\tau_{SRH}$ is Shockley-Read-Hall lifetime, $A$ is the radiative recombination parameter, and $n_i = \sqrt{N_c N_v}e^{-E_g/(2kT)}$ is the intrinsic charge carrier concentration in the semiconductor with the effective densities of states $N_c$ and $N_v$ in the conduction and valence bands, respectively, and the bandgap $E_g$.

As seen from (\ref{1}), the ideality factor equals 2. This value is not due to the recombination in the space-charge region, but due to the high excitation level and the rear surface contribution to the I-V curve.

The open-circuit voltage, $V_{OC}$, is obtained from (\ref{1}) by setting the current density to zero. Furthermore, by multiplying the current by voltage and setting the derivative $d(J_L(V)V)/dV$ to zero, one can determine the photogenerated voltage $V_m$ at maximal power. Substitution of this result into (\ref{1}) gives the corresponding current density $J_m$. Then, photoconversion efficiency under AM1.5 conditions can be found as  $\eta = J_m V_m/P_S$, where $P_S$ is the incident energy flux, equal to 0.1 W/cm$^2$ for AM1.5G conditions. Also, the I-V curve fill factor is given by $FF = J_m V_m/(J_{SC}V_{OC})$.

\begin{figure}[t!] 
\includegraphics[scale=0.3]{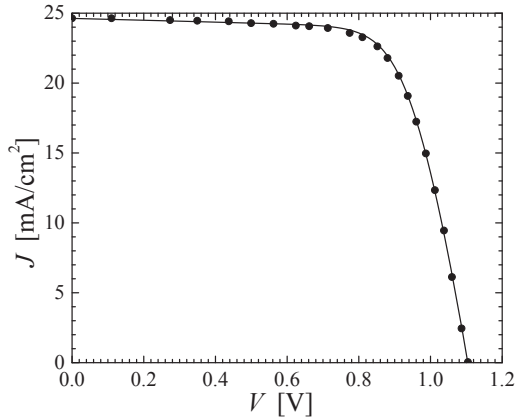}
\caption{Illuminated experimental I-V curve of a FAPbI$_3$-based SC from \cite{You14} (symbols) and theoretical fit (solid line). }
\label{fig1}
\end{figure}

Now, let us compare the theoretical results with the experimental findings from \cite{You14} obtained on a FAPbI$_3$ perovskite p-i-n SC. Shown in Fig.~\ref{fig1} is the experimental photogenerated current density vs. voltage curve, $J_L(V)$. The theoretical expression (\ref{1}) is in excellent agreement with the experimental data if the combination $J_0 \cong qn_i(d/\tau_{SRH} + S)$ is set to $1.15\cdot 10^{-8}$\,mA/cm$^2$, and the series and shunt resistance are taken to be $R_S = 4.5\,\Omega$, $R_{sh} =1.5\cdot10^3\,\Omega$. 
Using the data from table~\ref{table1}, we obtain the intrisic carrier density $n_i \approx 2.5\cdot10^5$\,cm$^{-3}$.
Given that the bandgap, $E_g$, in FAPbI$_3$ equals 1.55 eV (see \cite{You14}), this corresponds the densities of states product of $\sqrt{N_cN_v} \approx 3\cdot10^{18}$\,cm$^{-3}$. 

Note that the surface recombination velocity of 15\,cm/s turned out to be quite small, even though no special measures were taken to passivate the interfaces of the i-region with the n$^+$ and p$^+$ regions. Thus, the theory developed here agrees well with the experiment and gives the ideality factor equal to 2.

\begin{table*}[t!]
\centering
\caption{Experimental and theoretical parameters of perovskite SCs. The i-region thickness $d = 0.56\,\mu$m, as in \cite{You14}. Shockley-Read-Hall lifetime and radiative recombination parameter are taken from \cite{Wehrenfennig14}.}
\begin{tabular}{| l | l | l | l | l | l | l | l | l | l |}
\hline
 & $\tau_{SRH}$,\,s & $A$,\,cm$^3$/s & $S$,\,cm/s & $R_S,\,\Omega$ & $R_{sh},\,\Omega$ & $J_{SC}$,\,mA/cm$^2$ & $V_{OC}$,\,V & $\eta$,\,\% & $FF$,\,\%\\ \hline
\text{real} & $2\cdot10^{-7}$ & $9\cdot10^{-11}$ & 15 & 4.5 & $1.5\cdot10^3$ & 24.7 & 1.104 (1.104) & 19.7 (19.7) & 72.3 (72.4)\\
\hline
\text{limit} & $\infty$ & $9\cdot10^{-11}$ & 0 & 0 & $\infty$ & 26.2 & 1.22 & 29.2 & 90.8\\
\hline
\end{tabular}
\label{table1}
\end{table*}

\begin{figure}[t!] 
\includegraphics[scale=0.3]{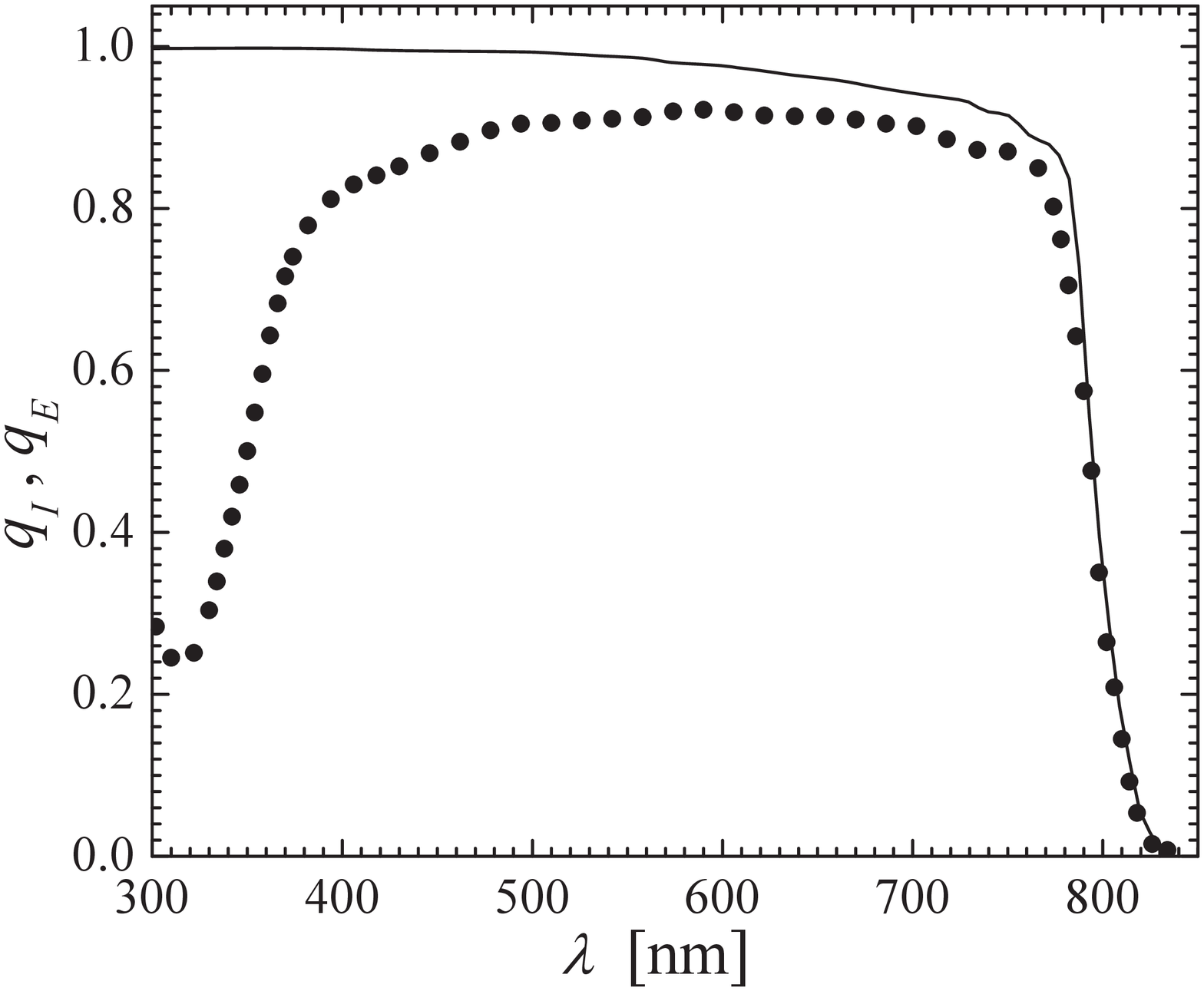}
\caption{Photocurrent external quantum efficiency in FAPbI$_3$ SCs (symbols) and theoretical internal photocurrent quantum efficiency obtained theoretically using (\ref{3}) (solid line).}
\label{fig2}
\end{figure}

Shown in Fig.~\ref{fig2} is the experimental wavelength dependence of external quantum efficiency of the photogenerated current, $q_E(\lambda)$, obtained in \cite{You14}. Knowledge of $q_E(\lambda)$ allows one to determine the short-circuit current as 
\begin{equation}
J_{SC} = q\int_{\lambda_0}^{\lambda_m}d\lambda\,I_{AM1.5}(\lambda)\,q_E(\lambda)\ ,
\label{2}
\end{equation}
where $\lambda_0$  is the short-wavelength absorption edge, $\lambda_m = 1240\,\text{nm}\cdot\text{eV}/E_g$ is the photoelectric threshold energy, and $I_{AM1.5}(\lambda)$ is the spectral density of radiation under AM1.5 conditions.

In the limiting case of no photocurrent losses due to light reflection and absorption in the layers that precede the perovskite layer, and due to the presence of contact grid (or semitransparent metal electrode), internal quantum efficiency, $q_I(\lambda)$, of a textured structure is given by the expression from \cite{Tiedje84}: 
\begin{equation}
q_I(\lambda) = \left(1 + \frac{1}{4\alpha(\lambda)dn_r^2}\right)^{-1}\ .
\label{3}
\end{equation}
Here, $\alpha$ is light absorption coefficient and $n_r$ is the perovskite refractive index. Substitution of (\ref{3}) into (\ref{2}) allows one to determine the limiting value of short-circuit current density.

Taking light absorption coefficient in FAPbI$_3$ from \cite{Kato16} and refractive index $n_r$ from \cite{Ndione16}, one can determined $q_I(\lambda)$, see the solid curve in Fig.\ref{fig2}. The theoretical $q_I(\lambda)$ curve also describes the experimental $q_E(\lambda)$ dependence near the absorption edge. This allows one to conclude that due to the non-planar location of grains that compose a perovskite, its surface is naturally textured, which leads to practically complete capture of the incident light due to reabsorption of reflected photons. This conclusion is supported by the values of $q_E(\lambda) \approx 92\%$ at the wavelength of about 600\,nm. If the SCs dealt with were of plane-parallel structure, then the light reflection coefficient, given by the Fresnel’s expression $(n_r - 1)^2/(n_r + 1)^2$, at $n_r = 2.06$ would be about 12\%, and the corresponding value of $q_E(\lambda)$ would be below 88\%, which is smaller than the experimentally measured value. 

Table~\ref{table1} contains the parameters used for comparison between theory and experiment, as well as the characteristics determined from this comparison. The second line contains the experimental characteristics of a real SC followed by the theoretical counterparts in the brackets. As seen from Table~\ref{table1}, the agreement between theory and experiment is very good. The third line of Table~\ref{table1} gives the limiting characteristics, which are obtained by neglecting all losses and Shockley-Read-Hall recombination. 

\begin{figure}[t!] 
\includegraphics[scale=0.3]{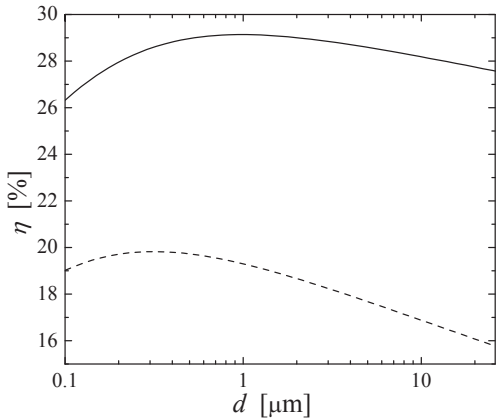}
\caption{Theoretical photoconversion efficiency as a function of i-region thickness obtained in the limiting case (solid line) and for realistic parameter values (dashed line). }
\label{fig3}
\end{figure}

Shown in Fig.~\ref{fig3} are the theoretical photoconversion efficiency vs. i-region thickness curves. The solid curve is built for the case of limiting parameter values, and the dashed curve is built for the realistic SC parameters. In both cases, the expression (\ref{3}) was used for the internal quantum efficiency. As seen in Fig.~\ref{fig3}, the respective curves develop a maximum, which is related to the fact that the photogenerated current increases with thickness, saturating at large thicknesses, whereas the photogenerated voltage decreases with thickness due to the increase of bulk recombination rate. The limiting efficiency at the maximum is about 29\%. In both cases, the maximum in rather broad, and is located at $d_{max} \approx 0.9\,\mu$m in the limiting case and at $0.33\,\mu$m in the realistic case. The experimentally used perovskite thicknesses are in the range 0.34 -- 0.6\,$\mu$m. 

Because some of the electron-hole pairs may form excitons with binding energy $E_x$ between 25 and 50\,meV \cite{Wehrenfennig14, Almansouri15, Chen16}, let us discuss their influence on the photoconversion efficiency. 

On the one hand, excitons have a positive effect on photoconversion efficiency, because they increase light absorption coefficient. On the other hand, their presence leads to the reduction of the open-circuit voltage and short-circuit current. It is possible to estimate the exciton effect on the open-circuit voltage by noticing that, according to \cite{Chen16},  $n_{i, ex}^2 = n_i^2e^{E_x/kT}$. This expression together with Eq.~(\ref{1}) allows one to estimate the reduction of $V_{OC}$, which turns out to be about 2 -- 4\,mV. Strictly speaking, in the case considered here, saturation current density $J_0$ is constant, because it is taken from experiment. This means that as $n_i$ increases, bulk recombination rate must decrease, i.e. bulk lifetime should grow. In \cite{Han16}, this lifetime in FAPbI$_3$ is estimated to be $\approx 5\cdot10^{-7}$\,s. If we take this value, then $n_i$ should be equal $5.7\cdot10^5$\,cm$^{-3}$, and exciton binding energy is estimated as 43.5\,meV. 

Let us now estimate the exciton influence on the short-circuit current. Because a part of photogenerated electron-hole pairs are bound into excitons, which do not conduct electricity, we should compare the exciton concentration, $n_x$, with the electron-hole pair concentration $\Delta n$. According to \cite{Kane93},
\begin{equation}
n_x = \frac{N_x\Delta n^2}{N_cN_v}e^{E_x/kT}\ ,
\label{4}
\end{equation}
where $N_c = \nu_n\left(2\pi m_n kT/h^3\right)^{3/2}$, $N_v = \nu_p\left(2\pi m_p kT/h^3\right)^{3/2}$, $N_x = \nu_x\left(2\pi (m_n + m_p) kT/h^3\right)^{3/2}$ are the effective densities of states in the conduction, valence, and exciton bands, respectively, and $\nu_{n, p, x}$ are the corresponding degeneracy factors. In direct-bandgap semiconductors, they are equal $\nu_n = \nu_p = 2$ and $\nu_x = 8$.

Estimates show that, depending on the electron and hole effective masses, $m_n$ and $m_p$, ranging from about 0.2 to 1 electron mass, and on the value of $\Delta n$, ranging between $10^{14}$ and $10^{15}$\,cm$^{-3}$, the value of $n_x$ turns out to be at least two orders of magnitude smaller than $\Delta n$. Thus, in the case considered, excitons reduce the short-circuit current by not more than 1\%.

Coming to our conclusions, the analysis of the illuminated I-V curve from \cite{You14} yields the ideality factor of 2. The same value is obtained within the theory described here at $\Delta n \gg n_0$ by taking into account the rear surface contribution. 

Based on the analysis of the experimentally measured external quantum efficiency of FAPbI$_3$-based SCs, an important conclusion can be made about surface texturing of these SCs, which results in capture and absorption of the reflected photons and essential increase of photoconversion efficiency. It is established that the theoretical efficiency vs. i-layer thickness curve, $\eta(d)$, has a maximum at $d$ in the $0.3 - 0.9\,\mu$m range. This maximum is due to the competition between the different dependences of short-circuit current and open-circuit voltage on the i-layer thickness. 

With the parameters given in Table~\ref{table1}, the theoretical and experimental I-V curves agree well with each other. Finally, it is established that in perovskite SCs, the surface recombination velocity has a rather low value even in the absence of passivation of the i-region boundaries.

\acknowledgments
M.E. is grateful to the Natural Sciences and Engineering Research Council of Canada (NSERC) and to the Research and Development Corporation of Newfoundland and Labrador (RDC) for financial support.

\end{document}